\documentclass[a4paper]{VisionStyle}
\usepackage{epsfig}

\begin{document}

\title{{\it Chandra} Observations of a New Supernova Remnant
 AX~J1843.8$-$0352/G28.6$-$0.1}

\author{Katsuji\,Koyama\inst{1}, Masaru\,Ueno\inst{1}, Aya\,Bamba\inst{1} \and
 Ken\,Ebisawa\inst{2}} 

\institute{
 Department of Physics, Graduate School of Science, Kyoto University, 
Sakyo-ku, Kyoto, 606-8502, Japan
\and 
{\it INTEGRAL} Science Data Center Chemin d'Ecogia 16 CH-1290 Versoix,
Switzerland}

\maketitle 

\begin{abstract}
\object{AX~J1843.8$-$0352} is a new X-ray SNR identified to
the radio complex \object{G28.6$-$0.1} with the $ASCA$ satellite.
$Chandra$ discovered two distinct components from this SNR: 
non-thermal and thin thermal X-ray emissions. 
The non-thermal component is fitted with 
a power-law spectrum of photon index 2.0.
The morphology is complicated,
but roughly an elliptical shape with a mean diameter of
about $7\arcmin$--$10\arcmin$. 
The east to south rims of the ellipse are associated
with the non-thermal radio sources C, F and G (Helfand et al. 1989).
The power-law slope of the radio spectrum can be smoothly connected
to that of X-rays with a break  at around the optical-IR band,
hence would be due to synchrotron X-rays
accelerated probably to $\geq$ 1 TeV at the shell of the SNR.

The thermal component near the southeast rim is a thin plasma
of about 0.8~keV temperature. 
It has the appearance of a "Tadpole" figure with a head of  
$30\arcsec\times 40\arcsec$-size and a tail of 30$\arcsec$-long.
Although this emission is associated with the west part of the radio source F,
the absorption is about two times larger than that of the non-thermal X-rays,
the bulk of the SNR emission.
Therefore, together with the peculiar morphology,
whether the thermal plasma is 
a part of the SNR or a background object is unclear.

\keywords{SNR: individual (AX~J1843.8$-$0352) --- X-rays: ISM --- Particle Acceleration}

\end{abstract}

\section{Introduction}

Supernovae (SNe) and their remnants (SNRs) play essential roles
for the structure and evolution of the Galaxy.
SNRs may be the major sources of the Galactic hot ($10^5$--$10^8$K) medium. 
The blast wave of SNe may compress the interstellar medium,
and trigger the successive star formations.
They produce and distribute heavy elements in the whole Galaxy
and even in the intergalactic space.

SNRs would also be the most plausible sites of cosmic ray production
near to the knee energy ($\sim10^{15.5}$~eV); 
supporting evidence is the detection of synchrotron X-rays
from the shell of some of the SNRs
(Koyama et al. 1995; 1997)
and the detection of inverse Compton gamma rays 
(Tanimori et al. 1998; Muraishi et al. 2000).   
High energy electrons up to about 1~TeV or even more are
produced possibly by the Fermi acceleration process. 
The energy loss rate of electrons, hence the synchrotron flux,
is proportional to $B^2$, while the  
gain (acceleration) rate
is proportional to $B$, where $B$ is strength of the magnetic field.
High energy electrons responsible to the synchrotron X-rays
can therefore exist in a shell of rather weak magnetic field
where the radio flux should be faint. 
Synchrotron X-rays are, in fact, found in radio faint shell-like SNRs.

SNRs may be the major origin of the Galactic ridge X-rays (GRXs).
The GRXs are diffuse X-rays extending along the Galactic inner disk.
The spectrum shows a thin thermal plasma of about $10^8$~K
having a prominent K-shell line from highly ionized irons.
There is mounting evidence that the GRXs are attributable to diffuse sources,
not an integrated emission of many point sources.
No cataloged diffuse source, however, can account the observed spectrum and 
flux of the GRXs.
Therefore the GRXs predict possible presence of either new X-ray SNRs or 
new category of X-ray sources.
Sensitive surveys with $ROSAT$ and $ASCA$ found several new diffuse X-rays,
which were later identified with radio faint SNRs.
At present, the number of Galactic SNRs in X-ray is less than a half of
that in radio (Green 2001).
It is plausible that more diffuse X-ray
structures can be found with highly sensitive 
X-ray surveys with $Chandra$.
The $Chandra$ deep exposure observations were performed
with these idea and found a diffuse structure (Ebisawa et al. 2001, also 
in this Proceedings),
which corresponds to the new SNR AX~J1843.8$-$0352 (Bamba et al. 2001).
This paper reports the results of the diffuse structure
and discusses possible implication 
on the origin of high energy cosmic rays and the GRXs.
 
\section{Observations and Data Reduction}

The $Chandra$ deep observations on the Galactic ridge were
performed on Feb. 24--25, 2000 (AO1, here Observation 1)
and on May 20, 2001 (AO2, here Observation 2), with respective
targeted positions of $(l, b)$ = (28$\fdg$450, -0$\fdg$200),
and (28$\fdg$550, -0$\fdg$029).
The satellite and instrument are described
by Weisskopf et al. (1996) and Garmire et al. (2000), respectively.   
The observations were made with the ACIS-I array,
which covers a $17\arcmin\times17\arcmin$ field. 
AX~J1843.8$-$0352 lies near the northwest edge and the east half of the ACIS-I
array of Obs. 1 and Obs. 2, respectively.
Data acquisition from the ACIS-I was made in the Timed-Exposure Faint mode
with the chip readout time of 3.24~s.
The data reduction and analysis were made using the $Chandra$ 
Interactive Analysis of Observations (CIAO) software version 2.1.
Using the Level 2 processed events provided by the pipeline
processing at the $Chandra$ X-ray Center,
we selected the ASCA grades 0, 2, 3, 4 and 6, as the X-ray events;
the other events, which are due to charged particles,
hot and flickering pixels, are removed. 
The effective exposure is  $\sim100$~ks for each observation.

\section{Results and Analyses}

Figure 1 is the mosaic ACIS-I image of Obs 1 and 2 in the 1.5--8~keV band
overlaid on  the 20~cm VLA contours.
The new $ASCA$ SNR, AX~J1843.8$-$0352 is clearly observed
at the radio complex G28.6$-$0.1.
Here and after we refer the radio source names (A-I) by Helfand et al. (1989).
 
\begin{figure}[ht]
\begin{center}
\epsfig{file=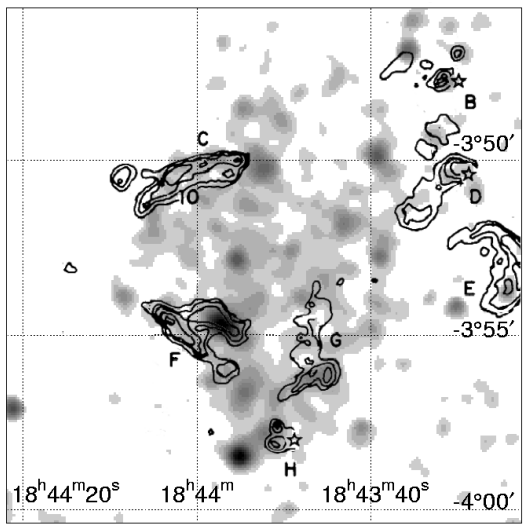, width=7cm}
\end{center}
\caption{
The ACIS-I image in the 1.5--8.0~keV band with logarithmic gray scale. 
The contours are the 20~cm radio fluxes (Figure 6 of Helfand et al.\ 1989). 
Large diffuse structure associated with the radio complex G28.6$-$0.1
is a new SNR AX~J1843.8$-$0352.
The brightest spot at the south is a discrete  source,
which may be unrelated object to the SNR.
A bright diffuse clump at the radio source F is a thin thermal plasma.}  
\label{kkoyama-B2_fig:fig1}
\end{figure}

The X-ray morphology is complicated with many clumps,
which are filled along and in the elliptical region of
the radio continuum complex.
The association with the X-ray clumps is found at C and F at the east rim. 
The south tail of the radio source G is also followed by X-rays.
The bright spot at the south is a discrete source,
which may be unrelated to the SNR.

\begin{figure}[ht]
\begin{center}
\epsfig{file=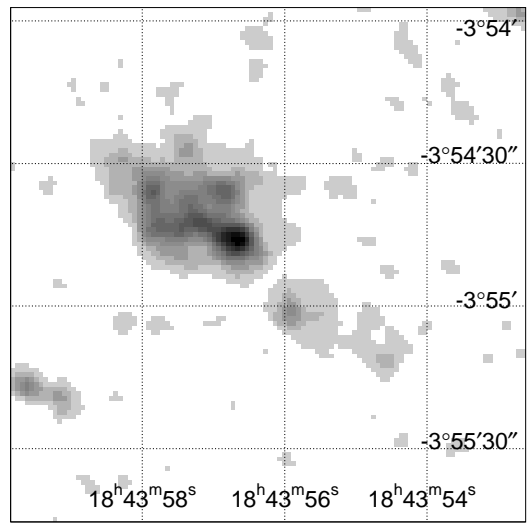, width=7cm}
\end{center}
\caption{
Closed-up view of the Tadpole in the
1.0--6.0~keV band with linear scale.
This shows a peculiar shape, an elliptical head 
with a jet-like tail of 30\arcsec-long toward the southeast.}
\label{kkoyama-B2_fig:fig2}
\end{figure}

Figure~2 is the closed-up view of the brightest clump
near the southeast rim in the 1.0--6.0~keV band.
We see the appearance a "Tadpole" with a $30\arcsec\times 40\arcsec$-size head
at around (18:43:57, -03:54:38)$_{J2000}$
and a tail of $30\arcsec$-long (here the Tadpole). 
The Tadpole shows clear association with the west tail of the radio 
source F.

The radio source E,
which has a non-thermal spectrum and a crescent shell in radio,
is the most secure candidate of a radio SNR.
However no X-ray is found near here.
Other radio sources B and D, possibly compact HII regions,  
are not associated with X-rays.

\begin{figure}[ht]
\begin{center}
\epsfig{file=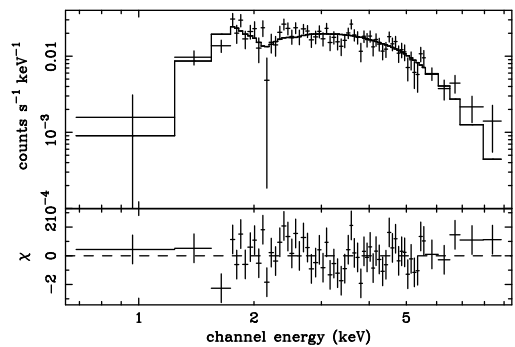, width=7cm}
\end{center}
\caption{
The X-ray spectrum of AX~J1843.8$-$0535 (crosses) with the best fit 
power-law model (solid histogram).}
\label{kkoyama-B2_fig:fig3}
\end{figure}

\begin{table}[bht]
  \caption{Power-law model of AX~J1843.8$-$0352}
  \label{kkoyama-B2_tab:tab1}
  \begin{center}
   \leavevmode
    \footnotesize
    \begin{tabular}[h]{ccc}
     \hline\hline \\[-5pt]
     Index & $N_{\rm H}$ & $F_{\rm X}^*$  \\
     $\alpha$  & $10^{22}$ Hcm$^{-2}$ & $10^{-12}$erg cm$^{-2}$s$^{-1}$ \\
     \hline \\[-5pt] 
     2.0 (1.6--2.3) & 3.5 (2.9--4.1) & 4.3  \\
     \hline\hline \\[-10pt] 
     \end{tabular}
\end{center}
$^*$Unabsorbed X-ray flux in the 0.7-10.0 keV band.\\
Parentheses are the 90\% error regions.
%\end{center}
\end{table}

We pick up many point sources using the "wavdetect" software
and  exclude all these point-like sources
for the analysis of the diffuse structures.
The integrated flux of the point sources in the AX~J1843.8$-$0352 
region is less than 10\% of the diffuse emission of the SNR,
hence the residual contamination, if any, is negligible. 
The spectrum of AX~J1843.8$-$0352 is made from an ellipse of 
10\arcmin (major axis) by 7\arcmin (minor axis),
where diffuse emission from the Tadpole is also excluded.  
The background spectrum is made from the same size of 
the ellipse at the same distance from the Galactic plane
as those of AX~J1843.8$-$0352, 
which is selected so as to properly subtract the GRXs 
(Koyama et al. 1986).

The background-subtracted spectrum, as is given in Figure 3,
shows no significant emission line, hence is fitted with a power-law  
model. The fitting is acceptable with the best-fit parameters given in Table~1,
and  the best-fit model shown in Figure 3 (the solid line).

\begin{figure}[ht]
\begin{center}
\epsfig{file=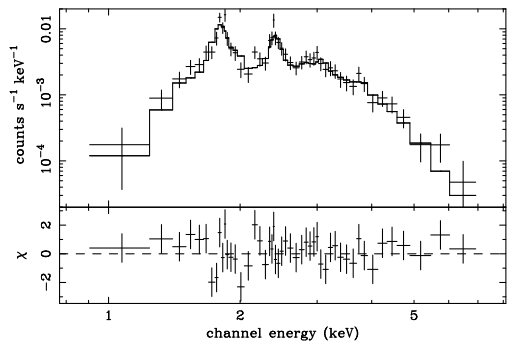, width=7cm}
\end{center}
\caption{
X-ray spectrum of the Tadpole (crosses) with the best fit 
NEI model (solid histogram).}
\label{kkoyama-B2_fig:fig4}
\end{figure}

 \begin{table}[bht]
  \caption{Thin thermal model of the Tadpole}
  \label{kkoyama-B2_tab:tab2}
  \begin{center}
   \leavevmode
   \footnotesize
   \begin{tabular}[h]{cccc}
   \hline\hline \\[-5pt] 
   $kT$ &$N_{\rm H}$ & $nt^\dagger$ & $F_{\rm X}^*$ \\
    keV & $10^{22}$Hcm$^{-2}$ & $10^{10}$s.cm$^{-3}$ & $10^{-11}$erg cm$^{-2}$s$^{-1}$ \\
   \hline \\ [-5pt] 
   0.84 & 6.4 &7.8 & 1.6 \\
   (0.73--0.95) & (5.8--7.2)  &(2.6--20) &  \\
   
\hline\hline \\[-10pt]
\end{tabular}
\end{center}
$^\dagger$Ionization parameter; $n$ and $t$ are the electron density
and elapse time of the plasma. \\
$^*$Unabsorbed X-ray flux in the 0.7--10.0 keV band.\\
Parentheses are the 90\% error regions.
%\end{center}
\end{table}

The X-ray spectrum of a peculiar source, the Tadpole is given in Figure 4, 
where the background taken from around the source is subtracted.
The Tadpole, in contrast to AX~J1843.8$-$0352,  shows clear emission 
lines at 1.85 keV and 2.36 keV,
which are equal or slightly lower than those from He-like Si and S.
We therefore fit the spectrum with an NEI (non equilibrium ionization)
plasma model of the solar abundances.
This model is acceptable with the best-fit parameters given in Table 2
and the best-fit model shown in Figure 4 (solid line).

\section{Discussion}

\subsection{AX~J1843.8$-$0352}

We found that the new SNR AX~J1843.8$-$0352 has many hard X-ray clumps,
which are along and in the elliptical region of
the radio complex G28.6$-$0.1.
The X-ray spectrum is well fitted with a power-law model
of photon index 2.0 (1.7-2.4), consistent with the $ASCA$ results.  
Since the $ASCA$ spectrum could not exclude a thin thermal model (NEI) 
of solar abundances,
we try this model to the $Chandra$ spectrum and find that the abundances
must be very low ($\leq$0.4 solar),
which does not favor a thin thermal origin.
The constraint on the AX~J1843.8$-$0352 model 
is thus improved due to the separation of the diffuse thermal clump
(the Tadpole) with the superior spatial resolution of $Chandra$.

The absorption ($N_{\rm H}$) of $ 3.5(\pm 0.6)\times 10^{22}$ Hcm$^{-2}$
is slightly larger than, but roughly similar to the $ASCA$ result.
Therefore, we adopt the source distance to be 7 kpc (Bamba et al, 2001).
The X-ray luminosity (0.7--10.0~keV) and the source size are then 
estimated to be $2.3\times 10^{34}$~erg~s$^{-1}$ and $14 \times20$ pc$^2$,
respectively.

The radio non-thermal sources C and F come near the east rim, 
although the association with X-rays is not good in
detail. The south part of the radio source
G is also associated with the X-ray rim.  
We therefore make a combined spectrum
from the radio to the X-ray bands and find that 
the radio (energy index = 0.5) and X-ray band spectra (1.0) 
are smoothly connected with a break near at the optical-IR band. 
The energy index of 0.5 in the radio band is explained
by synchrotron emission of high-energy electrons 
having a power-law distribution of index 2.0.
The larger energy index of 1.0 (photon index, 2.0) in the X-ray band
should be due to the synchrotron energy loss of higher energy electrons.
The wide band spectrum resembles those of SN~1006, G347.3$-$0.5
and RX~J0852.0$-$4622,
the well established shell-like SNRs as a site of high 
energy electrons (Koyama et al.1995; Koyama et al.1997;
Allen, Markwardt \& Petre 1999). 
We hence propose that AX~J1843.8$-$0352 is
another synchrotron X-ray dominant shell-like SNR.

\subsection{The Tadpole}  

The Tadpole has a thin thermal spectrum of 0.8~keV temperature 
with the solar abundances, which is typical of a young SNR.
Also the projected position is in the SNR AX~J1843.8$-$0352, hence
a naive scenario is that the Tadpole is a thermal component  of 
this SNR.
Then using the distance of 7~kpc,
the X-ray luminosity is estimated to be $\sim 10^{35}$~erg~s$^{-1}$ (0.7--10 keV). 
The physical size is  $\sim 0.9\times1.3$~pc$^2$ of head and 
$\sim$1.3~pc-long tail,
then the mean plasma density is  $\sim 10^{1.5} $ cm$^{-3}$.  
This value is extremely high compared with any other SNRs. 
The jet (tail)-like morphology resembles the hot plasma
associated with the jet source SS433.
We however see no central source at the head of the Tadpole.

Since the  absorption toward the Tadpole
($6.7\times 10^{22}$ Hcm$^{-2}$) is nearly two times larger than that 
of the host SNR AX~J1843.8$-$0352 ($3.5\times 10^{22}$ Hcm$^{-2}$),
and is more typical of those thorough the Galactic plane 
(Ebisawa et al. 2001), the possibility is not rejected that
the Tadpole is a background object at
either a far-side of the Scutum arm or out of our Galaxy.  

\subsection{The Origin of GRXs and Cosmic Rays}

Present discovery of the low surface brightness diffuse X-ray sources
both in the hard (AX~J1843.8$-$0352) and soft (the Tadpole) bands
may have important implication on the origin of GRXs and  cosmic rays.

The GXRs are prominent in the inner disk of the Galactic longitude
$l = \pm30\degr$, near the same position of\\ AX~J1843.8$-$0352 and the Tadpole. 
The GRXs have been found to exhibit two components (Kaneda et al. 1997).
One is a thin thermal plasma of 0.8~keV temperature
with prominent K-shell lines from He-like silicon and sulfur
(here, the soft component), similar to the Tadpole.
If the "Tadpole"-like sources are omnipresent,
they may largely contribute to the soft component of the GRXs.

The other component of GRXs is a 7--10~keV temperature plasma,
which emits strong K-shell line from highly ionized iron (the hard component).
AX~J1843.8$-$0352, if fitted with a thin thermal model,
shows about 6~keV temperature,
similar to that of the hard component of the GRXs.
The surface brightness of AX~J1843.8$-$0352 is about twice the GRXs
at this position.
Therefore, a part from the iron K-shell line,
significant contribution to the hard component (continuum emission) of GRXs
should be inferred.

We find that AX~J1843.8$-$0352 is a new candidate of synchrotron X-rays.
Since the radio and X-ray luminosity is the same order of SN~1006,
the contribution to the Galactic cosmic rays should also be the same.
The surface brightness of AX~J1843.8$-$0352 in radio and X-ray
is near or bellow the detection limit of the previous instruments.
We thus can predict many more
"AX~J1843.8$-$0352" in the Galactic plane,
which may ultimately account most of the cosmic rays in our Galaxy.

In order to see, whether the diffuse X-ray sources,
like AX~J1843.8$-$0352 and the Tadpole are omnipresent or not, 
and to approach the real origin of the GRXs and cosmic rays,
we encourage to perform further deep exposure observations
on the Galactic inner disk. 

\section{Summary}

\begin{enumerate}
\item   We separately detected diffuse hard X-rays from many discrete sources
 	in a new SNR AX~J1843.8$-$0352, which are associated with 
	the non-thermal radio complex\\ G28.6$-$0.1
	in the Galactic Scutum arm.\\
\item  The X-ray spectrum is fitted with  
    	a non-thermal power-law model of photon index 2.0.
	Together with the morphology of 14--20 pc diameter ellipse
	and the radio spectrum, 
	we conclude that AX~J1843.8$-$0352 is
	a synchrotron X-ray dominant SNR. \\
\item  We discovered a 0.8 keV plasma (the Tadpole) at the east rim of
	AX~J1843.8$-$0352 with a peculiar morphology of
	a elliptical head and jet-like tail. \\
\item  $N_{\rm H}$ absorption of the Tadpole is nearly two times larger
 	than that of the non-thermal SNR AX~J1843.8$-$0352.  
	Together with the peculiar morphology, whether the Tadpole is 
 	a thermal component
	of AX~J1843.8$-$0352 or unrelated background source is debatable.\\
\item   The new sources AX~J1843.8$-$0352 and 
	the Tadpole provide important implication on the origin of the
	the GRXs and the Galactic cosmic rays.\\
\end{enumerate}

\end{document}